\documentclass[%
reprint,
superscriptaddress,
floatfix,
amsmath,
amssymb,
aps,
notitlepage
]{revtex4-1}

\usepackage{placeins}

\usepackage[hidelinks=true]{hyperref}
\usepackage{xcolor}
\usepackage{threeparttable}
\usepackage{comment}
\usepackage{multirow}

\definecolor{greenish}{RGB}{108, 200, 105}
\definecolor{reddish}{RGB}{174,12,48}
\definecolor{blueish}{rgb}{0.12, 0.56, 1.0}
\definecolor{magenta}{RGB}{242, 80, 93}
\hypersetup{
  colorlinks   = true,
  urlcolor     = greenish,
  linkcolor    = blueish,
  citecolor    = magenta
}

\usepackage{amsmath}
\usepackage{setspace}
\usepackage{caption}
\usepackage{float}
\usepackage{siunitx}
\usepackage[utf8]{inputenc}
\usepackage{subcaption}
\usepackage{graphicx}
\usepackage{dcolumn}
\usepackage{bm}

\makeatletter
\newcommand*{\rom}[1]{\exp\!andafter\@slowromancap\romannumeral #1@}
\makeatother

\def\r{\mathbf{r}}

\def\rho{\varrho}
\def\eh{\hat{\mathbf{e}}}

\def\L{\mathcal{L}}
\def\argmin{\mathrm{argmin}}

\begin{document}
\title{Geometry and design of popup structures}
\author{Jay Jayeshbhai Chavda}
\affiliation{Department of Applied Mechanics \& Biomedical Engineering, IIT Madras, Chennai, TN 600036.}
\author{S  Ganga  Prasath}
\email{sgangaprasath@smail.iitm.ac.in}
\affiliation{Department of Applied Mechanics \& Biomedical Engineering, IIT Madras, Chennai, TN 600036.}

\begin{abstract}
Origami and Kirigami, the famous Japanese art forms of paper folding and cutting, have inspired the design of novel materials \& structures utilizing their geometry. In this article, we explore the geometry of the lesser known popup art, which uses the facilities of both origami and kirigami via appropriately positioned folds and cuts. The simplest popup-unit resembles a four-bar mechanism, whose cut-fold pattern can be arranged on a sheet of paper to produce different shapes upon deployment. Each unit has three parameters associated with the length and height of the cut, as well as the width of the fold. We define the mean and Gaussian curvature of the popup structure via the discrete surface connecting the fold vertices and develop a geometric description of the structure. Using these definitions, we arrive at a design pipeline that identifies the cut-fold pattern required to create popup structure of prescribed shape which we test in experiments. By introducing splay to the rectangular unit-cell, a single cut-fold pattern is shown to take multiple shapes along the trajectory of deployment, making possible transitions from negative to positive curvature surfaces in a single structure. We demonstrate application directions for these structures in drag-reduction, packaging, and architectural fa\c{c}ades.
\end{abstract}

\maketitle
\section{\label{Introduction}Introduction}
\noindent The geometry of a thin sheet of paper can be tuned by introducing cuts and folds at appropriate locations. Origami and Kirigami, the Japanese art forms of folding and cutting a sheet of paper, has inspired the design of soft robots~\cite{Misseroni2024,DangKirigami,rafsanjani2018kirigamiSkins, prashanth2024kirigamiRobots, jin2020kirigamiInflatables}, biomedical devices, prosthetics~\cite{brooks2022kirigamiBiodesign}, and architectural structures~\cite{li2021shapeMorphingEnvelopes, melancon2021multistable} that are capable of large-scale shape reconfiguration from a flat sheet with diverse mechanical behavior. These art forms have also inspired the design of grid-shells~\cite{chaudhary2021totimorphic,panetta2019xshells,segall2024fabricTessellation,pillwein2021elasticGeodesicGrids} that borrow the features of origami and kirigami in the form of tessellated structures. Such structures have been of interest to engineers, artists, and architects alike because they are scalable, modular, and are often deployed by a single actuating degree-of-freedom, attributes arising out of their geometric nature and patterning of repeating units. Several approaches to program shape in these materials and structures have been introduced~\cite{dudte2016programming,choi2019programming, choi2021compact,jiang2020freeform,celli2018shapeMorphingSheets, redoutey2021popupDome}, but once fabricated, these structures transform to a single shape or operate within a limited curvature regime. This limitation arises due to the capability of a single cut/fold pattern or assembly to transform to a single shape and the stiff isometric/geometric constraints set by the inextensibility of thin sheets or rigidity that hinder their mobility.

In this article, we look at the geometry of a popup structure~\cite{petzallUllagami} made out of a single sheet of paper by cutting and folding at appropriate locations. Similar to earlier approaches~\cite{dudte2016programming, choi2019programming, choi2021compact}, we understand the geometry of a popup structure made from assembling several units as a function of the deployment. We find that popup structures, unlike origami- and kirigami- based designs, offer a large space of possibility to program shape while not compromising on the deployability and the rigidity of the structures. This arises because of the combination of cuts and folds, that roughly translate to introducing floppiness and rigidity in the sheet. In order to fabricate popup structures that are scalable to a large number of cuts and folds as well as to different desired shapes, we develop an optimization based pipeline that is tested against experiments. By introducing splay to each popup unit, we further show that a single cut-fold pattern can transform to multiple shapes along the deployment trajectory. We demonstrate this by fabricating a structure that transforms from positive Gaussian curvature to negative during deployment. Using our design framework, we fabricate popup structures that are used to modify airfoil profiles, conform to shapes for packaging, and as architectural fa\c{c}ades.

\begin{figure*}[t]
    \centering
    \includegraphics[width=\textwidth]{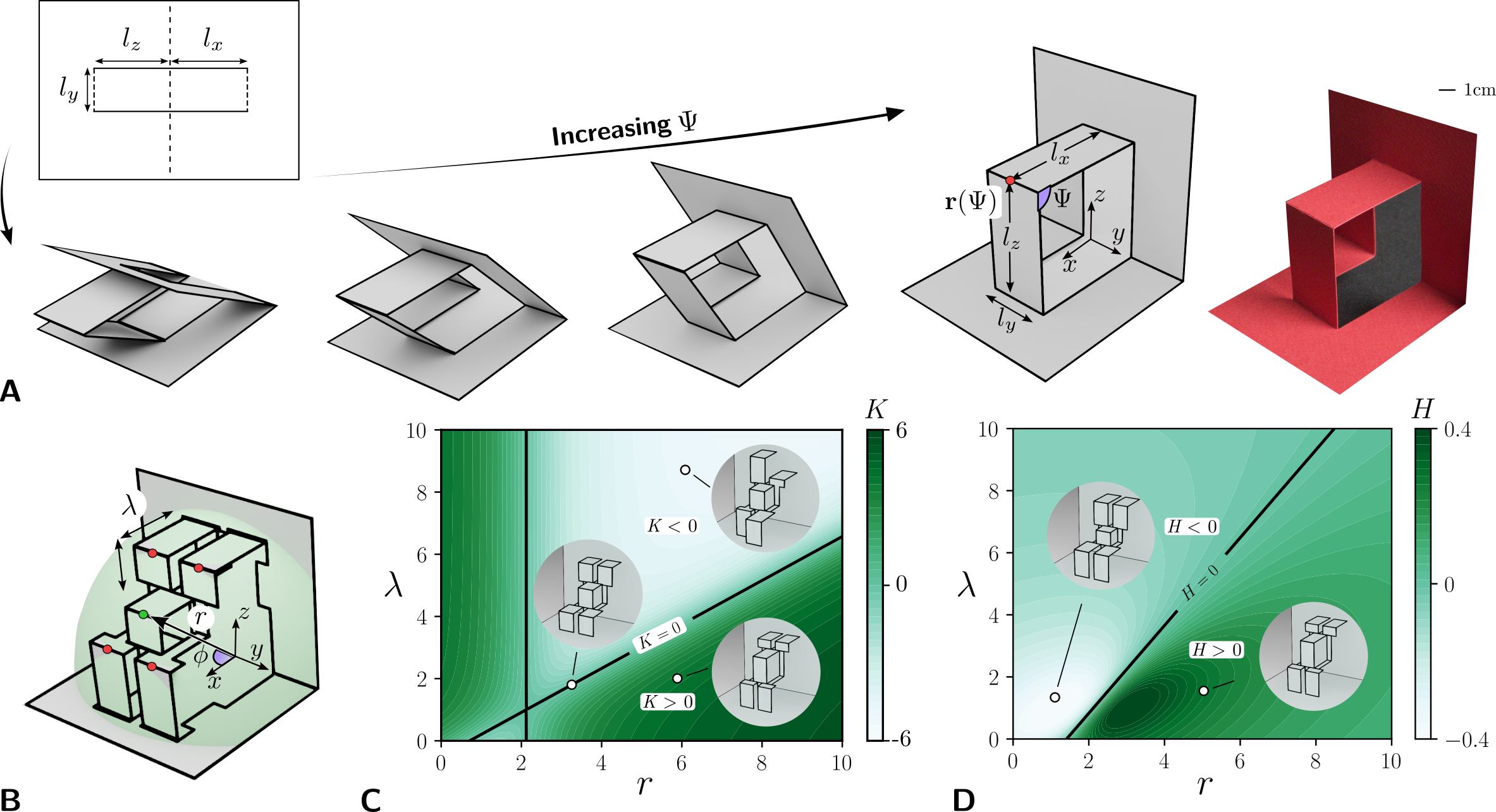}
        \caption{\textsf{\textbf{Geometry of popup structures.}}
        (A) A single popup unit is made from a flat sheet by two straight cuts and three parallel folds (cut-fold pattern shown at top left), parameterized by lengths $l_x, l_z$, and width $l_y$. Increasing actuation angle $\Psi$ (from 0 to $\pi/2$) drives deployment through intermediate states to the final 3D configuration. The fold vertex $\r(\Psi)$ is located at the center of the width. The image on the right in red is a popup unit from experiments. (B) Assembly of five popup units used to define curvature of a popup structure with a central unit parameterized by ($r$, $\phi$); $r, \phi$ are the distance and the angle made by the vector joining the fold vertex of central unit-cell with the origin, while $\lambda$ is scaling factor that changes the lengths of of opposite pair of unit-cells. For $\lambda > 1$, the diagonal units are larger than the rest and vice-versa for $\lambda < 1$. (C) Gaussian curvature $K(r,\lambda)$ of the five-unit-cell assembly is computed by triangulating the fold vertices of the units and using the Gauss-Bonnet theorem (described in detail in SI Sec.~\ref{Gaussian-and-mean}). The $K=0$ solid line divides positive ($K>0$, convex) and negative ($K<0$, saddle/hyperbolic) regions, enabling different curvature regimes by tuning $(r, \lambda)$. (D) Mean curvature $H(r,\lambda)$ of the popup assembly computed using the discrete Laplace-Beltrami operator (described in main text). Locus of $\lambda, r$ corresponding to minimal surface, $H=0$ divides locally convex and concave regions of the deployed structure. $K(r,\lambda)$ and $H(r,\lambda)$ describe the geometry of popup structures with fixed widths.}
    \label{fig:fig1}
\end{figure*}

\section{\label{sec:geometry} Results}
\subsection{Geometry of popup units}
\begin{figure*}[t]
    \centering
    \includegraphics[width=\textwidth]{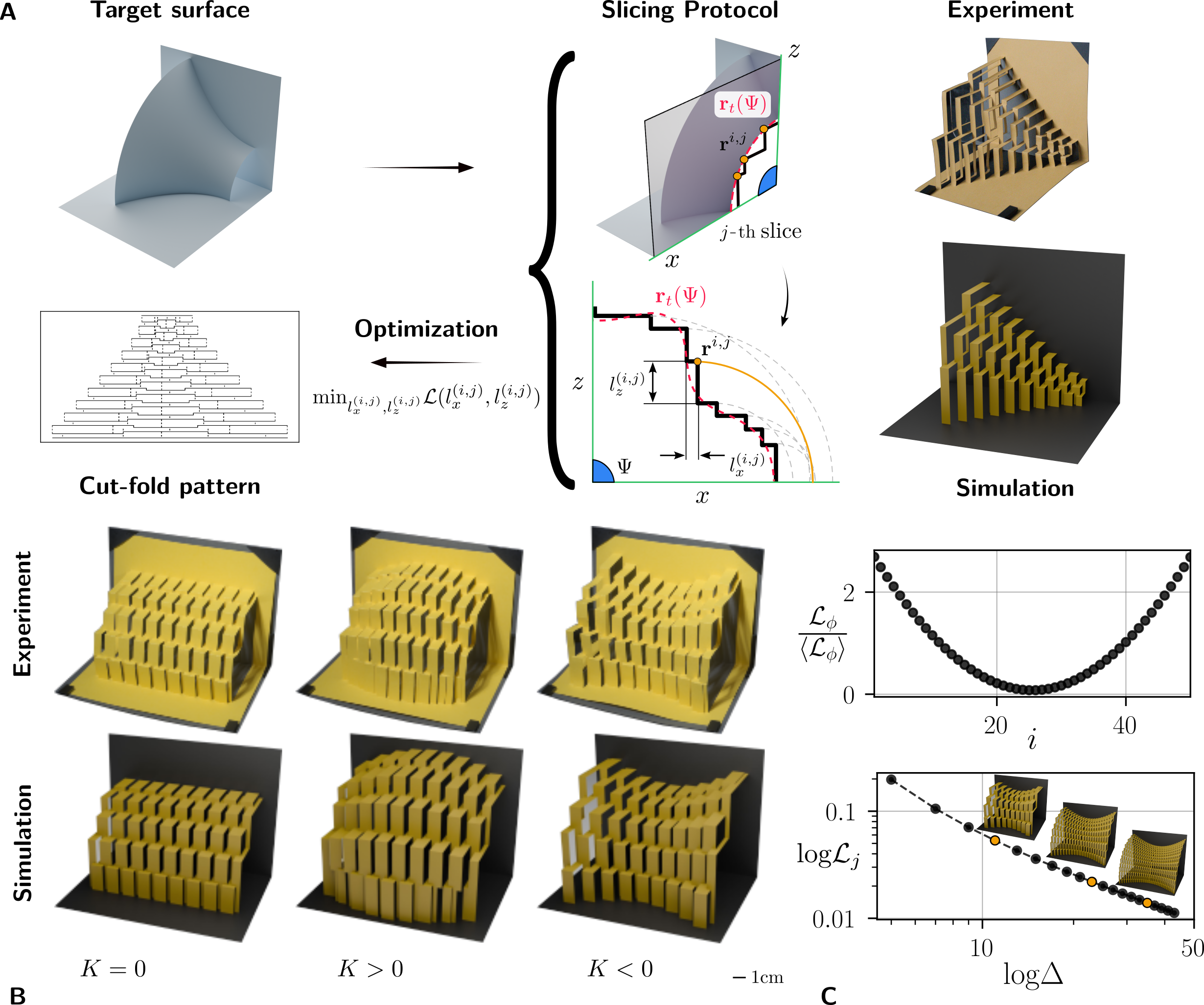}
    \caption{\textsf{\textbf{Pipeline for popup structure design.}}
    (A) A popup structure with uniform width is designed to transform to a target 3D surface by first slicing the target surface $\r_t(\Psi=\pi/2)$ along the transverse direction at equidistant locations to get 2D curves. Fold vertices $\r^{i, j}$ on these curves, parameterized by ($l_x^{i,j}$, $l_z^{i,j}$) -- the length and height of cut of $i$-th popup unit of $j$-th slice, are obtained by minimizing $\L_j[\{l_x^{i,j},\,l_z^{i,j}\}]$ (see main text and SI Sec.~\ref{subsec:slice-opt} for details). The cost function, $\L_j$ ensures neighboring popup units are similar in dimension and are all of approximately the same size. This minimization procedure is subject to isometric and topological constraints, resulting in a physically realizable popup structure. The solution $\{l_x^{i,j*},\,l_z^{i,j*}\}$ obtained from this procedure is converted to cut-fold pattern that is fed into cutting machine for prototyping. Figures on the right show experimental and simulation models of the target surface developed using this pipeline. (B) Physical prototypes (top row) and simulation results (bottom row) demonstrating popup structures with zero $(K=0)$, positive $(K>0)$, and negative $(K<0)$ Gaussian curvature (see SI Sec.~\ref{sec:branching}, Table~\ref{tab:expdetails} for details). (C) (Top): Normalized azimuthal error density $\mathcal{L}_\phi / \langle \mathcal{L}_\phi \rangle$ vs unit-cell index $i$, evaluated from the solution $\{l_x^{i,j*},\,l_z^{i,j*}\}$ for a quarter-circle, which is useful to capture all axi-symmetric target shapes. We see that minimum error occurs at the center fold, $i=(N+1)/2$ due to symmetry constraints that ensure match with a circular target and the error increases towards the edges due to the effect of curvature (see main text for details). (Bottom): Log-log plot of the loss function $\mathcal{L}_j$ vs number density $\Delta = L/N$, demonstrating convergence and homogeneous structure with increasing density of popup units.}
    \label{fig:fig2}
\end{figure*}
\noindent The fundamental building block of our popup structure is a unit composed of two straight cuts (of lengths $l_x, l_z$) and three parallel folds (of width $l_y$) on a flat sheet of paper (shown in Fig.~\ref{fig:fig1}A). The fold vertex $\r(\Psi)$ (seen as the red dot in Fig.~\ref{fig:fig1}A, is defined as the midpoint along the fold width) of the unit and moves along a trajectory given by,
\begin{align}
\mathbf{r}(\Psi) = \bigl[l_x + l_z \cos(\Psi)\bigr] \eh_x + l_z \sin(\Psi) \eh_z, \label{eq:unit_kinematics}
\end{align}
where $\Psi$ is the angle of deployment. This unit resembles a four-bar linkage in which the panels of the sheet act as rigid links with a single actuating degree-of-freedom $\Psi$ that varies between $0$ and $\pi$. These units are tiled to create a popup structure whose geometry is described by considering five neighboring units (shown in Fig.~\ref{fig:fig1}B). The fold vertices (red and green dots in Fig.~\ref{fig:fig1}B) of these units can be connected by triangles to form a closed polygonal loop on the deployed surface (see SI Sec.~\ref{sub:gaussian_curvature} for further details). The effective Gaussian curvature of the popup structure, $K(r,\lambda)$ is computed from this triangulation using the Gauss--Bonnet theorem, $K(r,\lambda) = 2\pi - \sum_{i=1}^{4} \alpha_i(r,\lambda)$, where $\alpha_i(r,\lambda)$ are the interior angles at the central vertex. The variable $r$ controls the radial distance of the outer fold vertices from the central one and therefore sets the overall size of the configuration of five-cells on the surface. The scaling variable $\lambda$, stretches the diagonal pair of units and changes the vertex angles in a controlled way (see Fig.~\ref{fig:fig1}B). Figure~\ref{fig:fig1}C shows the map of $K(r,\lambda)$ and we find that along two black lines $r = (1+2\lambda)/\sqrt{2}, 3/\sqrt{2}$ the Gaussian curvature $K=0$. The presence of the $K=0$ lines highlights that by moving in the $(r,\lambda)$-plane, the same five-cell building block can realize a negative, zero or positive Gaussian curvature. The effective mean curvature $H(r,\lambda)$ of the popup structure at the central vertex can be computed using the discrete Laplace-Beltrami operator to get
\begin{align*}
H(r,\lambda) &= \frac{1}{2A}\sum_{(i,j)} \bigl(\cot\gamma_{ij} + \cot\epsilon_{ij}\bigr)\,\bigl(\r_i - \r_j\bigr)\cdot \hat{\mathbf{n}},
\end{align*}
where the sum runs over the edges $(i,j)$ incident on the vertex, $\gamma_{ij}$ and $\epsilon_{ij}$ are the angles opposite to the edge $(i,j)$ in the two adjacent faces, $A$ is the associated mixed (Voronoi) area, $\r_i$ are the vertex positions, and $\hat{\mathbf{n}}$ is the local unit normal. Figure~\ref{fig:fig1}D shows the contour plot of the mean curvature $H(r,\lambda)$, with the black curve marking the locus $H=0$ given by $r=(1/\sqrt2)(2+\lambda)$. For any chosen pair of $(r,\lambda)$ we can now compute the local $K(r,\lambda)$ and $H(r,\lambda)$ of the deployed popup structure (see SI Fig.~\ref{fig:SI_fig5_newloss} for a generalization of this). These definitions, as we shall see, help in programming shapes into popup structures by choosing appropriate combinations of $(r,\lambda)$. An important feature we see from Figs.~\ref{fig:fig1}C, D is that several combinations of $(r,\lambda)$ can result in the same $K(r,\lambda)$ and $H(r,\lambda)$, making popup structures with a single degree-of-freedom amenable to varied design choices.

\subsection{Programming shape in cut-fold pattern}
\noindent In order to program the shape of a popup structure that takes a desired three-dimensional shape upon deployment, we develop a design pipeline which outputs the cut-fold pattern that can then be used for fabrication in experiments (see SI Sec.~\ref{sec:branching} for experimental details). The design procedure (shown schematically in Fig.~\ref{fig:fig2}A) starts with a target surface $\r_t(x, y)$ that is sliced along its width. The series of fold vertices $\r^{i,j}(\{l_x^{i, j}, l_z^{i, j}, l_y^{j}\}; \Psi)$ described by their lengths and heights $\{l_x^{i, j}, l_z^{i, j}\}$ at width $l_y^{j}$ are chosen in the curve that results from the slicing process. The values of $(l_x^{i, j},l_z^{i, j})$ of the $i$-th popup unit are obtained by solving an optimization problem that minimizes the shape error between the target surface $\r_t(x, y)$ and the fold vertices $\r^{i,j}(\Psi)$ while ensuring that each unit is as close to a square as possible. Below we describe this procedure and demonstrate the resulting cut-fold patterns in experiments.\\

\noindent \textit{Slicing and optimization}: Given a target surface $\r_t(x,y) = \bigl(x,\, y,\, r_t(x,y)\bigr)$, slicing them with planes $y=y_j$ results in curves $\r_t^j(x) = \bigl(x,\, y_j,\, r_t(x,y_j)\bigr)$. The fold vertices on the $j$-th slice are represented using the length and height of the cuts, $l_x^{i,j}, \; l_z^{i,j}$ (see Fig.~\ref{fig:fig2}A, right) and for a given combination of $\{l_x^{i,j}, \; l_z^{i,j}\}$, unit-cell kinematics follows $\r^{i,j}(\Psi)=\bigl[(\sum_{k=1}^{i}l_x^{k,j}) + (L-\sum_{k=1}^{i}l_z^{k,j}) \cos(\Psi)\bigr] \eh_x$ $+ (L-\sum_{k=1}^{i}l_z^{k,j}) \sin(\Psi) \eh_z$. These fold vertices $\r^{i,j}(\Psi)$ at $\Psi = \pi/2$ of the $N$ popup units per curve are then placed on the curve $\r_t^j(x)$. Although infinite choices of $\{l_x^{i,j}, \; l_z^{i,j}\}$ may satisfy the positivity constraints, $l_x^{i,j}, \; l_z^{i,j} > 0$ and isometric constraints, $\sum_i l_x^{i,j} = \sum_i l_z^{i,j} = L$ (see SI Sec.~\ref{subsec:slice-opt} for further details), we seek a smooth  distribution of cells that simultaneously matches the target curve $\r_t^j(x)$ while minimizing the loss,
\begin{align}\label{eq:optimization}
\L_j[\{l_x^{i,j},\,l_z^{i,j}\}]
&= \sum_{i=1}^{N-1} \bigl\|l_x^{i+1,j} - l_x^{i,j}\bigr\|^{2}
 + \bigl\|l_z^{i+1,j} - l_z^{i,j}\bigr\|^{2}  \notag\\
&\quad + \sum_{i=1}^{N} \Bigl\|l_x^{i,j} - \Delta \Bigr\|^{2}
 + \Bigl\|l_z^{i,j} - \Delta \Bigr\|^{2}.
\end{align}
This loss while minimization procedure is applied, is supplemented with geometric constraints $\r_t^{i,j}\!\left({\pi}/{2}\right) - \mathbf{r}^{i,j}\!\left({\pi}/{2}\right) = 0$, and topological constraints $\mathbf{r}^{i+1,j}\!\left({\pi}/{2}\right) - \mathbf{r}^{i,j}\!\left({\pi}/{2}\right) \geq 0$ (see SI Sec.~\ref{SI_augmented_loss} for an alternative equivalent form of loss and SI Sec.~\ref{subsec:slice-opt} for more details about constraints). The first two terms in $\mathcal{L}_j$ penalize rapid changes in $l_x^{i,j}$ and $l_z^{i,j}$ along the slice, ensuring that neighboring units have similar sizes and avoid highly distorted cells. The last two terms demand a uniform unit-cell size, $\Delta = 1/N$ which prevents vanishing or excessively large units. Geometric constraints enforce that each deployed vertex lies on the prescribed target point $\r_t^{i,j}$, while the topological inequality maintains a consistent ordering of vertices along the slice and suppresses self-intersections or overlaps. This constrained optimization (depicted schematically in Fig.~\ref{fig:fig2}A), produces a sequence of unit-cell parameters, $\{l_x^{i,j*},\,l_z^{i,j*}\} = \argmin\ \L_j[\{l_x^{i,j},\,l_z^{i,j}\}]$, which can then be converted to a cut-fold pattern for fabrication.

For axisymmetric surfaces, the curves along the slices are quarter circles with different radius at $\Psi=\pi/2$. The behavior of the solution from the optimization procedure for such surfaces can be visualized by the normalized azimuthal error density, $\mathcal{L}_\phi/\langle\mathcal{L}_\phi\rangle$ (shown in Fig.~\ref{fig:fig2}C (top)) along the unit-cell index in the azimuthal coordinates. This error, defined as $\mathcal{L}_\phi = (\phi^{j} - j \Delta\phi)^2$ with $\phi^j$ being the angle made by the fold vertex with the $x$-axis, $\Delta\phi = \pi/(2N)$, is minimum in the center and increases monotonically on either side towards the edges. This is due to the fact that the fold vertices that lie on the circular target curve cannot arise from $l_x^{i, j} = l_z^{i, j}$ and we expect smoothly varying deviations away from the center. Furthermore, the loss $\mathcal{L}_j$ (shown in Fig.~\ref{fig:fig2}C (bottom)) for the popup structure, as expected, decreases with an increase in density of the unit-cells along the surface.\\

\noindent \textit{Visualization and fabrication:} The optimized solution across all slices yields a discrete set of fold vertices $\mathbf{r}^{i,j}(\Psi)$, which serve as anchor points for constructing the popup structure by introducing rectangular sections whose edges intersect at these vertices (ref.~SI Sec.~\ref{subsec:mesh-cut} for details). This assembly assumes: $(a)$ a tree-like topology for the branched structure obtained from the fold vertices, $(b)$ the area of the rectangular sections sum up to the area of the sheet, $(c)$ the topology of the assembly is preserved during deployment as we change $\Psi$.

\begin{figure*}[t]
    \centering
    \includegraphics[width=\textwidth]{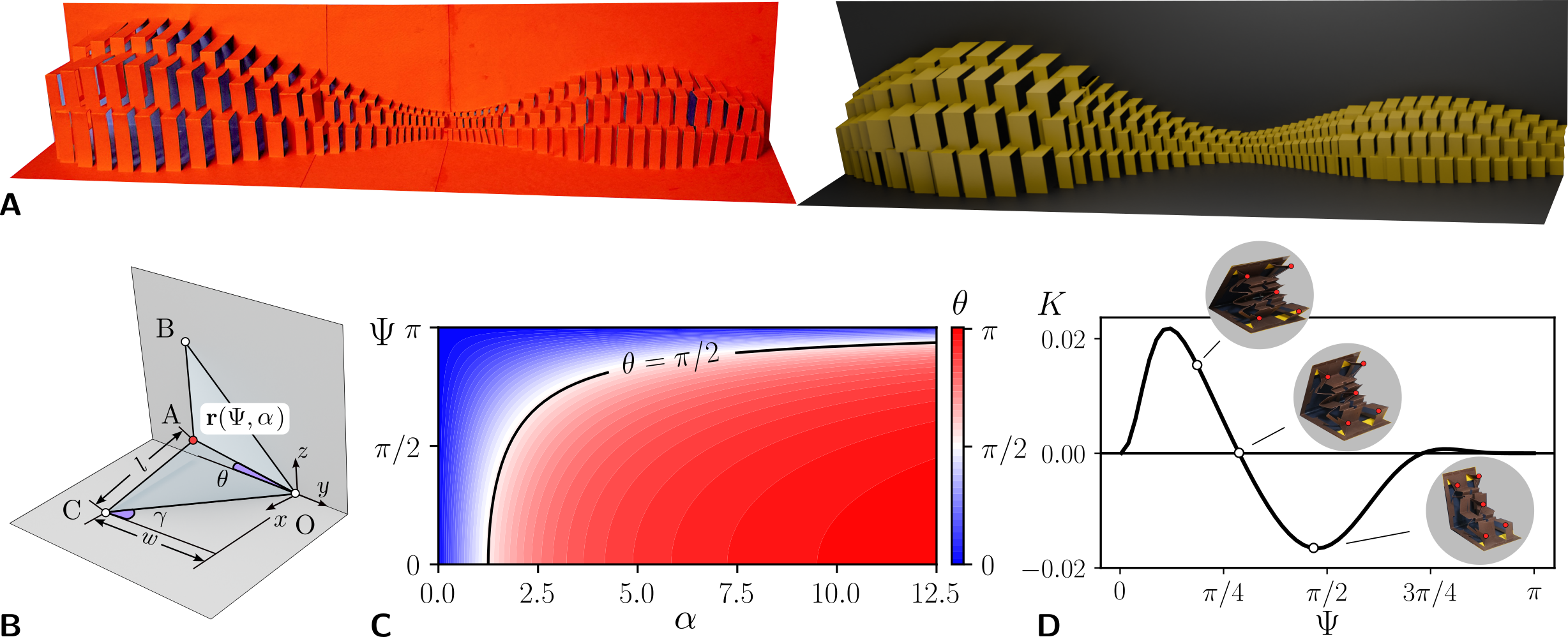}
    \caption{\textsf{\textbf{Non-uniform popup and multi-state popup with splayed units.}} (A) Popup structure containing all three gaussian curvature regimes ($K>0, K<0, K=0$) designed with popups whose widths vary along the longitudinal direction (see SI Sec.~\ref{subsec:width_variation}, Table~\ref{tab:expdetails} for details). The algorithm to compute the cut-fold pattern (described in the main text) accommodates complex surfaces with spatially varying curvature. (B) Schematic of a splayed unit-cell described by the vertex position, $\r(\Psi,\alpha)$ which is a function of the actuation angle, $\Psi$ and the slope, $\alpha$. (C) Contour plot of the orientation of the fold in the splayed unit, $\theta(\Psi,\alpha)$ in the $(\Psi,\alpha)$-plane (see SI Sec.~\ref{SIsplayed} for details). The level set curve corresponding $\theta=\pi/2$ (black curve) separates configurations in which increasing $\alpha$/varying $\Psi$ results in higher/lower vertex positions. Crossing this boundary enables transitions in gaussian curvature during deployment as the actuation angle, $\Psi$ varies. (D) Gaussian curvature, $K$ vs deployment angle, $\Psi$ of the splayed popup structure undergoing transition from $K>0$ state at $\Psi = 0.16\pi$ to $K=0$ at $\Psi = 0.28\pi$ to $K<0$ at $\Psi = 0.47\pi$. Inset shows the shape of the structure in experiments.}
    \label{fig:fig3}
\end{figure*}

For experiments, the solution $\{l_x^{i,j*},\,l_z^{i,j*}\}$ is converted to a cut-fold pattern as a vector design that can be directly used in the Cricut Maker~3 cutter (as shown in Fig.~\ref{fig:fig2}A). Our algorithm for creating these patterns uses straight solid lines to encode cuts and dashed lines to encode folds in the form of micro-cuts (see SI Video 1 for an example). Figure~\ref{fig:fig2}B shows examples of popup structures fabricated using this procedure corresponding to target surfaces with $K=0$, $K>0$, and $K<0$ (see SI Video 2 for the deployment of these structures). This approach also allows us to create structures with varying unit-cell widths that can encode diverse curvatures as shown in Fig.~\ref{fig:fig3}A (ref. SI Sec.~\ref{subsec:width_variation} for details). 

\subsection{Multi-state structures with splayed units}
\noindent We have seen so far that the 3 parameters, $l^{i, j}_x, l^{i, j}_y, l^{i, j}_z$, associated with each popup unit provide us with the capability to program a shape in these structures whose geometry evolves with deployment. However, the shape of the entire structure during its deployment is determined by the cut-fold pattern, limiting the possible shapes that a given pattern can be programmed to only one. We now show that even with a single cut-fold pattern, a popup structure can be programmed to take multiple shapes by introducing an additional variable at the unit-cell level that evolves with the actuation angle, $\Psi$. This additional variable that modifies the geometry of the unit-cell is a \textit{splay} introduced through two-fold symmetric folds that make an angle to the center fold (shown in Fig.~\ref{fig:fig3}B). The slope of this fold $\alpha = \tan \gamma$, is the splay associated with the unit and when $\alpha = 0$ the geometry reduces to the rectangular unit we have explored earlier. For each unit, $\theta$ is the orientation of the fold and depends on the actuation angle $\Psi$ and the slope $\alpha$ by the relation (see SI Sec.~\ref{SIsplayed} for details),
\begin{align*}
\theta(\Psi, \alpha)
= \tan^{-1}\!\left[
\frac{2\alpha \cos\!\left(\Psi/2\right)}
     {1 - \alpha^2 \cos\!\left(\Psi/2\right)^2}
\right].
\end{align*}
\noindent The fold orientation $\theta$ and the vertex position $\r(\Psi, \alpha)$ now depend on the actuation angle $\Psi$ and the splay $\alpha$, rather than $\Psi$ alone as in the rectangular case. We find that a variety of combinations of $\alpha, \Psi$ result in $\theta=\pi/2$ (see Fig.~\ref{fig:fig3}C), corresponding to the maximum out-of-plane height on the locus of the fold vertex $\r(\Psi, \alpha)$ and away from this curve in either $\Psi$ or $\alpha$ causes the vertex height to decrease monotonically. Thus, for a given actuation angle $\Psi$, we can find a unique $\alpha$ at which the splayed unit reaches a maximum height and choose values of $\alpha$ for neighboring units away $\theta=\pi/2$ to design popup structure that change their geometry upon deployment.

\begin{figure*}[htbp]
    \centering
    \includegraphics[width=\textwidth]{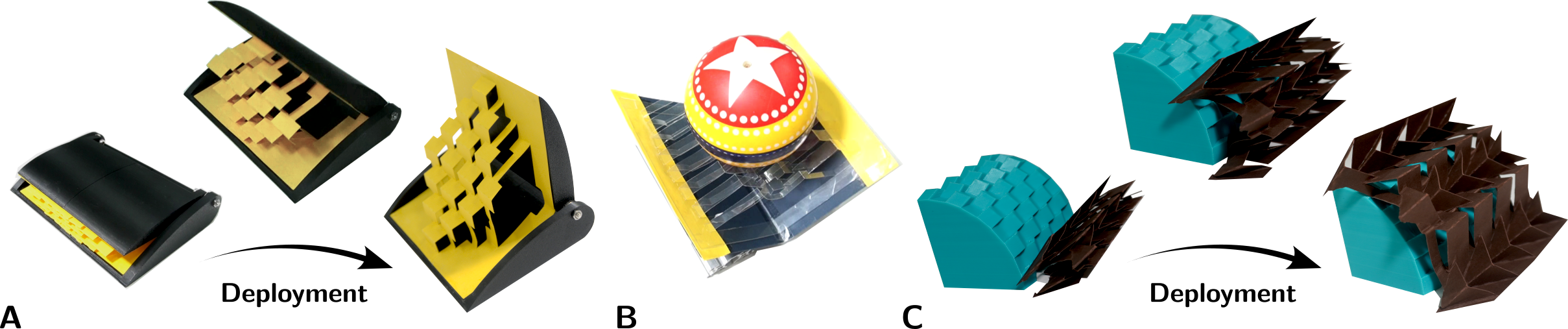}
        \caption{\textbf{\textsf{Potential applications of popup structures.}} (A) Sequence of snapshots showing the deployment of a popup structure attached to airfoil for possible drag reduction applications. (B) Popup structure made of a single sheet of Polypropylene sheet conforming around a sphere for packaging. (C) Snapshots showing a popup structure with splayed units acting as facade that controls the amount of light around the architecture.}
    \label{fig:fig4}
\end{figure*}

We use this new variable $\alpha$ to construct structures that can take multiple shapes from arrays of splayed units along the deployment trajectory. Consider a structure created with a cut-fold pattern of identical $l_x^{i,j}, l_z^{i,j}$, but the splay $\alpha^{i,j}$ is chosen independently for each unit. For such a  structure, along the deployment trajectory, all units rotate by the same $\Psi$, yet their fold vertices reach different heights according to their assigned $\alpha^{i,j}$. By appropriately selecting the values of $\alpha^{i,j}$ across the cut-fold pattern, the principal curvatures of this surface can be modified during deployment: in particular, we realize a popup structure that transforms from a shape with positive gaussian curvature, $K>0$ to a negative gaussian curvature, $K<0$ (see Fig.~\ref{fig:fig3}D and SI Sec.~\ref{SIsplayed} for more details).

\subsection{Potential applications}

\noindent Popup structures offer a wide range of potential applications, of which we explore three in this article. Airfoils with graded flaps can be deployed using popup mechanisms such that the opening of the flap results in interaction between the tail-vortices and the popup structure, affecting the drag of the wings (shown in Fig.~\ref{fig:fig4}A). By tuning the geometry of the units, the overall drag can be modulated depending on the operational requirements. They can also be used in the soft-packaging of products so that the geometry of the popup structure engulfs the object to be transported as shown in Fig.~\ref{fig:fig4}B. In architecture, deployable and reconfigurable fa\c{c}ades made of popup structures (see Fig.~\ref{fig:fig4}C) can take advantage of these patterns to create lightweight building skins that regulate daylight and capture solar energy while remaining compact and flat-packed.

\section{Discussion and Conclusions}
In this article, we have described the geometry of popup structures created from a single sheet of paper. In addition, we have developed a design framework to program desired shapes of these popup structures onto a cut-fold pattern. These structures, like other geometric materials, can be deployed from a flat state simply by bringing the ends of the sheet together. The programmability of these popup structures arises out of the three parameters associated with each popup unit, the length and height of the cuts and width of the fold. By introducing a modest modification to the fundamental unit in the form of splay, we show that a given cut-fold pattern can take multiple shapes along the deployment trajectory. Using this additional facility, we have demonstrated that a single cut-fold pattern can transform to surfaces with positive and negative gaussian curvature during deployment.

Although we have not explored non-straight cuts and folds in this article, we believe that such explorations open up a variety of possibilities worthy of a separation investigation which we will proceed in the future. Further, we have focused only on the geometric aspect of popup structures in this article, the effects of mechanics at the level of popup units still remain -- this is expected to play an important role when the cut-fold pattern is designed in a manner that it can result in stress-localization at the fold vertices or ends of cuts during deployment. Nevertheless, popup structures borrow the capabilities of origami and kirigami while structurally lying in the space of grid-shells making them geometrically unique. Designing these structures using our pipeline provides new avenues for developing materials with function that go beyond the realm of art~\cite{petzallUllagami}.\\

\section{\label{Methods}Appendix}
\noindent \textit{Experiments}: Our paper models of popup structures were made using 220~gsm sheets (Colorissimi by Favini). The cut-fold patterns were created with a Cricut Maker~3 equipped with the fine-point blade tool. Micro-cuts were inscribed on the sheets, and they act as guides for the formation of the folds during deployment. For large popup structures (such as the multi-unit structure shown in Fig.~\ref{fig:fig3}A), the global layout was first partitioned into multiple segments, each segment was cut separately, and the resulting pieces were then carefully aligned and glued along their common boundaries to reconstruct the final continuous surface from a single pattern.\\

\noindent \textit{Numerics}: All numerical optimization and inverse-design computations to minimize objective functions subject to geometric and topological constraints on the deployed fold vertex positions were performed in \textsc{Matlab} using the \texttt{fmincon} routine in the optimization toolbox. To convert the optimized unit-cell parameters into machine-readable cut files, the planar cut and fold lines were assembled into vector graphics and exported as SVG. The same exported graphs were imported into 3D rendering software \textsc{Blender} to generate visualizations of the final prototype geometries and to understand the deployment kinematics.\\

\noindent \textbf{Acknowledgments}: We thank IIT Madras and ANRF-ECRG/2024/003341/ENS for partial financial support. We also thank the INTERFACE lab members for their interactions. The data and the SI Videos are available at \href{https://github.com/sgangaprasath/Popup2026/}{Github}.
\bibliographystyle{apsrev4-1}
\bibliography{Popup}

\large
\widetext
\clearpage
\onecolumngrid
\begin{center}
\textbf{\large Supplemental Information for \\[.2cm]
``Geometry and design of popup structures''}\\[.2cm]
Jay Jayeshbhai Chavda$^{1}$, S  Ganga  Prasath$^{1}$\\[.1cm]
{\small \itshape ${}^1$Department of Applied Mechanics \& Biomedical Engineering, IIT Madras, Chennai, TN 600036.\\
}
\end{center}
\setcounter{equation}{0}
\setcounter{figure}{0}
\setcounter{table}{0}
\setcounter{page}{1}
\setcounter{section}{0}
\makeatletter
\setstretch{1.5}

\renewcommand{\thetable}{S\arabic{table}}%
\renewcommand{\thesection}{S\arabic{section}}
\renewcommand{\thesubsection}{SS\arabic{subsection}}
\renewcommand{\theequation}{S\arabic{equation}}
\renewcommand{\thefigure}{S\arabic{figure}}

\noindent \href{https://drive.google.com/file/d/19lfWjJQHDijkFalhwmXztKjd-UYR2iMT/view?usp=sharing}{\textbf{SI Video 1: Fabrication of a popup structure.}} Video demonstrating the fabrication process of the popup structure shown in Fig.~\ref{fig:fig2}A. We show the cutting process of the patterned sheet using the Cricut Maker~3. Then we see the folding process where the paper sheet is folded with the help of tweezers to form the final popup configuration.\\

\noindent \href{https://drive.google.com/file/d/1_n7_e4i7s_u0tAeZFS4V31970WHYwtcv/view?usp=sharing}{\textbf{SI Video 2: Deployment of popup structures.}} Video showing the experimental deployment of the popup structures shown in Fig.~\ref{fig:fig2}B with $K=0, >0, <0$. These structures were fabricated using the cut-fold pattern obtained via the design pipeline.

\section{Curvature of five-unit-cell assembly}\label{Gaussian-and-mean}
\noindent In this section, we derive the expressions for the Gaussian and mean curvature of the five-unit-cell popup assembly. The gaussian curvature, $K(r, \lambda)$ is computed using the Gauss-Bonnet theorem for the discrete vertices formed by the fold vertices of the assembly. The mean curvature, $H(r, \lambda)$ on the other hand, is calculated using the discrete Laplace-Beltrami operator \cite{Meyer2003Discrete}.
\subsection{Gaussian curvature}\label{sub:gaussian_curvature}
\begin{figure}[htbp]
\centering
\includegraphics[width=\textwidth]{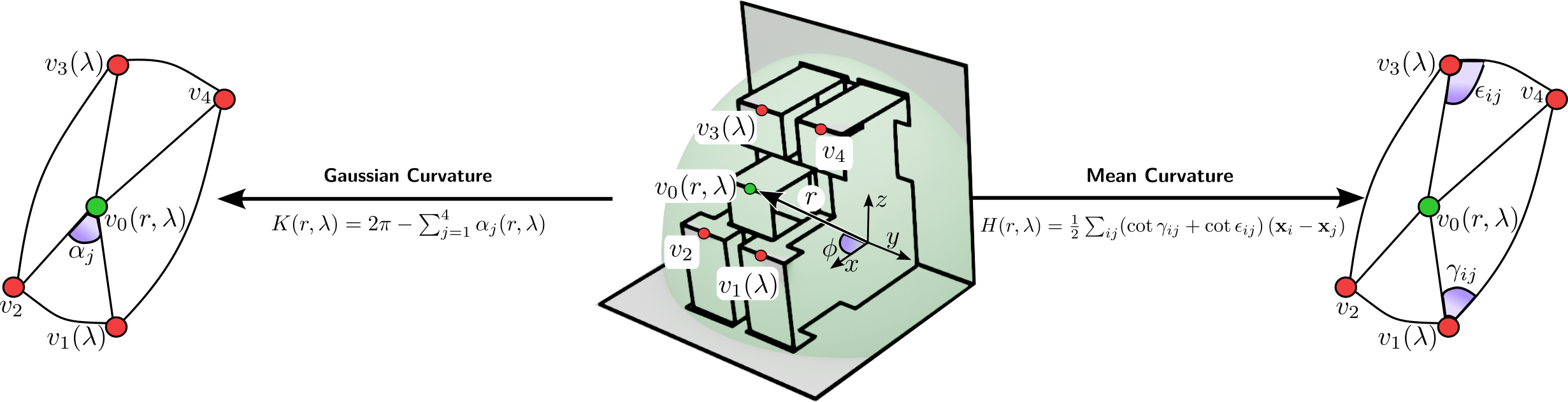}
\caption{
\textbf{\textsf{Illustration of curvature computation using fold vertices $\mathbf{r}^{i,j}$ in the five-unit-cell assembly.}} {Left:} Gaussian curvature $K$ is calculated using discrete Gauss--Bonnet: the angle defect $2\pi - \sum_{k=1}^4 \alpha_k$ in the star-shaped neighborhood (green central vertex, red boundary vertices) yields $K = 2\pi - \sum \alpha_k$. 
{Right:} Mean curvature $H$ is calculated using the cotangent Laplace operator: the mean-curvature normal $\mathbf{H}_0$ at the vertex is computed from edge vectors weighted by cotangents of opposite angles in adjacent triangles, then projected onto the surface normal $\mathbf{n}_i$ to obtain the signed scalar $H_0 = (\mathbf{H}_0 \cdot \mathbf{n}_i)/2$.}
\label{fig:curvature_illustration}
\end{figure}

\noindent \textit{Gaussian curvature definition:} For an interior fold vertex $v_0(r,\lambda)$, shown in Fig.~\ref{fig:curvature_illustration}, we consider the neighborhood formed by the four triangles that share this vertex and are bounded by its four neighboring vertices $v_1(\lambda)$, $v_2$, $v_3(\lambda)$, $v_4$ (see Fig.~\ref{fig:curvature_illustration}, left). This local patch is topologically equivalent to a disk and, therefore, has the Euler characteristic $\chi(S)=1$. The boundary polygon connecting the neighboring vertices follows straight edges in 3D space, which we treat as geodesics on the surface metric, so the geodesic curvature term along the boundary vanishes. Under these assumptions, the Gauss-Bonnet theorem reduces to $K = 2\pi - \sum_{j=1}^{4} \alpha_j(r,\lambda)$, where $\alpha_j$ are the four angles at the central vertex $v_0(r,\lambda)$, measured in the intrinsic surface metric. In practice, each $\alpha_j$ is computed from the 3D coordinates of the central vertex $v_0(r,\lambda)$ and its two neighboring vertices $v_i$ ($i=2,4$) and $v_j(\lambda)$ ($j=1,3$) using the law of cosines,
\begin{align*}
\alpha_j(r,\lambda)
&= \arccos\!\left(
\frac{\|v_0(r,\lambda)-v_i(\lambda)\|^2 + \|v_0(r,\lambda)-v_j\|^2 - \|v_i-v_j\|^2}
{2\,\|v_0(r,\lambda)-v_i(\lambda)\|\,\|v_0(r,\lambda)-v_j\|}
\right),
\end{align*}
and the resulting signed value $K$ is assigned to that vertex. Positive values $K>0$ correspond to dome-like regions, $K=0$ to locally flat configurations, and $K<0$ to saddle-like regions. Evaluating $K$ over all interior fold vertices yields the Gaussian-curvature contour plot $K(r,\lambda)$ shown in Fig.~\ref{fig:fig1}C.\\[5pt]

\noindent \textit{Mean curvature definition:} Mean curvature at the fold vertices of the five-unit-cell assembly is computed using the standard discrete mean-curvature normal operator defined on a triangulated surface. The deployed configuration is first represented as a triangular mesh whose vertices coincide with the fold vertices and whose faces approximate the local geometry around each unit-cell. At every vertex $v_i$, an area-weighted unit normal $\mathbf{n}_i$ is obtained by averaging the normals of all incident triangles and normalizing, so that $\mathbf{n}_i$ encodes the local orientation of the surface at that point. For any interior edge $(v_0(r,\lambda),v_i)$ shared by two triangles $(v_0(r,\lambda),v_i,v_{i-1}(\lambda))$ and $(v_0(r,\lambda),v_i,v_{i+1}(\lambda))$, we denote by  $\gamma_{ij}$ and $\epsilon_{ij}$ the angles at the opposite vertices $v_i$ and $v_{i+1}(\lambda))$, respectively. The cotangents of these angles are,
\begin{align*}
\cot\gamma_{ij}
&=
\frac{(\r_{v_0} - \r_{v_{i-1}})\cdot(\r_{v_i} - \r_{v_{i-1}})}
     {\|(\r_{v_0} - \r_{v_{i-1}})\times(\r_{v_i} - \r_{v_{i-1}})\|},
\\
\cot\epsilon_{ij}
&=
\frac{(\r_{v_0} - \r_{v_{i+1}})\cdot(\r_{v_i} - \r_{v_{i+1}})}
     {\|(\r_{v_0} - \r_{v_{i+1}})\times(\r_{v_i} - \r_{v_{i+1}})\|},
\end{align*}
where $\r_k$ denotes the position of the vertex $k$~\cite{Meyer2003Discrete}. These cotangent factors serve as weights that measure the degree to which the surface bends across the edge $(v_0(r,\lambda),v_i)$. Using these weights, the discrete mean-curvature normal vector at the vertex $v_0(r,\lambda)$ is defined as,
\begin{align}
\mathbf{H}_0
&=
\frac{1}{2A_i}
\sum_{j\in\mathcal{N}(i)}
\bigl(\cot\gamma_{ij} + \cot\epsilon_{ij}\bigr)\,
\bigl(\mathbf{r}_0 - \mathbf{r}_i\bigr),
\end{align}
where $\mathbf{r}_{0,i}$ is the position of $v_{0,i}$, $\mathcal{N}(i)$ is the
set of vertices connected to $v_0$ by an edge, and $A_i$ is a reference
area associated with $v_0$, taken here as the sum of one third of the areas
of all triangles incident on $v_0$~\cite{Meyer2003Discrete}. The signed mean curvature at the vertex $v_i$ is then obtained by projecting the
mean-curvature normal onto the vertex normal, $H_0 =\mathbf{H}_0\cdot\mathbf{n}_i/2$, so that $H_0>0$ and $H_0<0$ correspond to locally convex and concave bending with respect to the chosen orientation of $\mathbf{n}_i$~\cite{Meyer2003Discrete}. Evaluating this expression for all fold vertices and interpolating the resulting values in polar coordinates ($r$, $\phi$); $r, \phi$ are the distance and angle made by the vector that joins the fold vertex of the central unit-cell with the origin, which produces the signed mean–curvature field $H(r,\phi)$ shown in Fig.~\ref{fig:curvature_illustration}(for fixed $\phi=\pi/4$), where the contour $H=0$ delineates the transition between regions of net convex and concave bending over the five-unit-cell assembly.

\section{Branching protocol to generate cut-fold pattern}\label{sec:branching}

The branching protocol converts a prescribed target surface into a manufacturable cut-fold pattern by combining slice-wise constrained optimization with a tree-like branching assembly. The overall procedure consists of three computational stages: (1) generation of slice locations; (2) nonlinear optimization of fold vertices on each slice; (3) branching-based construction of a connected three-dimensional strip network which is subsequently converted into a triangulated mesh and a fabrication-ready cut-fold pattern.

\begin{figure}[htbp]
\centering
\includegraphics[width=\textwidth]{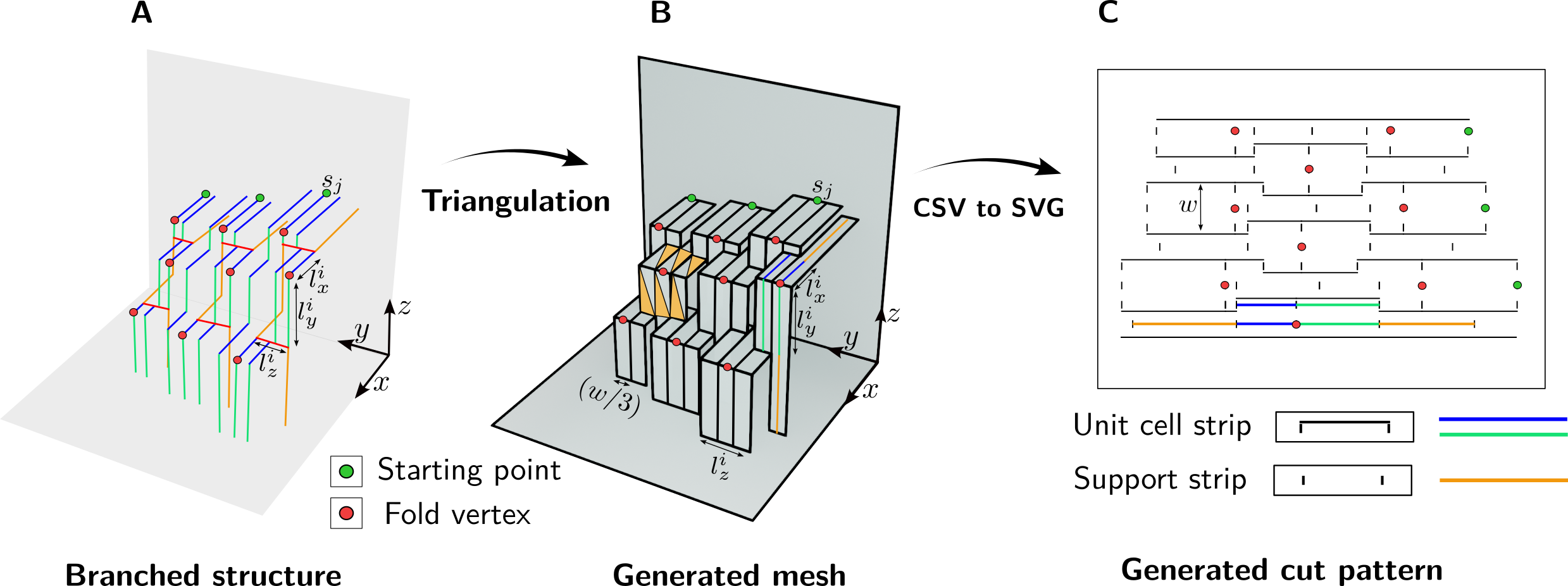}
\caption{\textbf{\textsf{Computational pipeline generating cut-fold pattern from optimized solution.}} (A) Optimized 3D branched structure defined by fold vertices (red points). Blue segments correspond to unit-cell panels oriented along the $x$ direction with length $l_x^{i}$, while green segments correspond to panels oriented along the $z$ direction with length $l_z^{i}$. Red segments indicate the separation between consecutive unit-cells by $l_y^{i}$, and yellow segments represent support strips. (B) Mesh generation: each structural segment is thickened into a finite-width rectangular panel and triangulated into two facets. (C) Final manufacturable cut-fold pattern (CSV to SVG): solid lines represent cuts and dashed lines represent folds; support strips require only folds for connectivity.}
\label{fig:Branching_structure}
\end{figure}

\subsection{Slicing and parameter definitions}

The target surface $\r_t(\Psi)$ is sampled by a set of planes aligned along the longitudinal direction (ref. Fig.~\ref{fig:fig2}A). Each plane intersects the target surface and results in a one-dimensional target curve $\r^j_t$ that must be approximated by a chain of fold vertices $\r^{i, j}$. The number of fold vertices per slice is denoted by $N$ and this controls the discretization resolution. Larger values of $N$ yield smoother approximations (ref. Fig.~\ref{fig:fig2}C) of the target curve but increase the computational cost. Slices are indexed by $j$, and the in-plane fold vertices along a slice are indexed by $i=1,\dots,N$. The distance of the slice $j$ along the $y$-axis is $s_j$ and the associated height of the fold vertex is $z(s_j)\in[0,1]$. In the case where the width of each unit is fixed for the entire popup structure, the unit-cell width is set to $l_y = w = (1/N)$, and the slice locations are spaced by a distance $jw$. More generally, our framework supports non-uniform slice-widths represented by vector $[w_1,\dots,w_{N_s}]$, where $w_j$ defines the local unit-cell width at slice $j$. In this case, the location of the fold vertices in the slice (indicated by green dots in Fig.~\ref{fig:Branching_structure}) is given by $s_j = \sum_{k=1}^{j-1} w_k,\,j=1,\dots,N_s$. For each slice $j$, the intersection of the target surface with the slice plane defines a target curve $\r_t = ( x, y_j, r_t(y_j, z(s_j))$, which prescribes the desired relationship between the optimized in-plane coordinate $l_x$ and height coordinate $l_z$ along that slice.

\subsection{Optimization of fold vertices, $\r^{i, j}$}\label{subsec:slice-opt}
\noindent For each slice $j$, the fold vertices $\r^{i, j}$ are represented by two unknown sequences, $\{l_x^{i,j},\,l_z^{i,j}\}_{i=1}^{N}$, which define the discrete chain approximating the target curve. The fold vertices are computed by solving a constrained optimization problem whose objective function includes smoothness cost through first differences and uniformity cost to prevent highly irregular segment distributions. Specifically, the loss function $\L_j$ is given by,
\[
\mathcal{L}_j
=
\sum_{i=1}^{N-1} \big(l_x^{i+1,j}-l_x^{i,j}\big)^{2}
+
\big(l_z^{i+1,j}-l_z^{i,j}\big)^{2}
+
\sum_{i=1}^{N} \Big(l_x^{i,j}-\frac{1}{N}\Big)^{2}
+
\Big(l_z^{i,j}-\frac{1}{N}\Big)^{2}.
\]
In addition, manufacturability and geometric feasibility are enforced through boundary, topology, and box constraints. The endpoints of the vertex chain are fixed as, $\mathbf{r}^{1,j}(\pi/2)=0\hat{\mathbf{e}}_x + 0\hat{\mathbf{e}}_z,\quad \mathbf{r}^{N,j}(\pi/2)=1\hat{\mathbf{e}}_x + 1\hat{\mathbf{e}}_z$, and topology constraints are imposed along the chain, $\mathbf{r}^{i+1,j}\!\left({\pi}/{2}\right) - \mathbf{r}^{i,j}\!\left({\pi}/{2}\right) \geq 0 $ where $i=1,\dots,N-1$. Furthermore, all vertices are restricted to lie within the admissible slice region using constraints $0\leq||\mathbf{r}^{i,j}(\pi/2)||\leq \sqrt{2}$. Finally, each fold vertex is constrained to lie exactly on the target curve, $\r_t^j$ corresponding to the $j$-th slice. The resulting constrained optimization problem is solved independently for each slice using \textsc{Matlab}'s \texttt{fmincon}.

\subsection{Topology of 3D popup structures}

After slice-wise optimization, a connected three-dimensional popup structure is assembled by constructing branches that traverse neighboring slices. The branching protocol produces a sparse tree-like topology while ensuring that the resulting sheet remains connected and manufacturable. Branches are initiated only from alternate slices, $j = 1, 3, 5, \dots$ which avoids dense cuts. For each starting slice $j$, the branch begins at the lowest fold vertex of that slice, and growns in discrete steps. Branch growth alternates between forward and backward motion in the slice index and for odd levels $i$, the algorithm advances from slice $j$ to $j+1$, while for even levels it returns from slice $j+1$ to $j$. A feasibility condition is enforced such that if the step leads to a negative in-plane increment (corresponding to backward motion in the $x$ direction), the branch is terminated. This condition prevents the construction of invalid configurations that would lead to overlap or self-intersection (shown in Fig.~\ref{fig:SIfig3}A). For each valid step, two orthogonal families of line segments are generated:
$(i)$ segments aligned with the in-plane direction (corresponding to unit-cell panels parallel to the $x$-axis shown by \textit{blue lines} in Fig.~\ref{fig:Branching_structure}A), and
$(ii)$ segments aligned with the vertical direction (corresponding to unit-cell panels parallel to the $z$-axis shown by \textit{green lines} in Fig.~\ref{fig:Branching_structure}A). These line segments represent the center lines of the unit-cell panels in the deployed configuration. In addition, separation segments (shown by \textit{red lines} in Fig.~\ref{fig:Branching_structure}A) are inserted between consecutive levels to represent the stacking distance between slices. Finally, support strips (shown by \textit{yellow lines} in Fig.~\ref{fig:Branching_structure}A)) are introduced to stabilize the structure and preserve connectivity between levels. These strips are represented as additional line segments positioned between unit-cell panels. The support-strip width is chosen proportional to the local slice width, ensuring that the amount of retained material scales consistently with the unit-cell discretization. All generated segments are stored in a manner that each entry contains the start and end coordinates of the segment, its associated length, and a tag indicating whether the segment corresponds to a unit-cell panel, a connector element, or a support strip. The complete line network is exported as a CSV file which forms the main geometric output of the branching construction stage.

\subsection{Mesh generation and deployment simulation}\label{subsec:mesh-cut}

\noindent The CSV file is converted into a triangulated surface mesh for visualization and simulation. Each line segment is interpreted as the center line of a rectangular panel. A finite panel width is assigned based on the local slice width, and each segment is thickened into a quadrilateral patch (refer Fig.~\ref{fig:Branching_structure}B). Each quadrilateral is then triangulated into two triangular facets, producing a watertight triangulated representation suitable for STL export. To visualize deployment, the structure is evaluated over a discrete set of deployment angles,
\[
\Psi_k = \frac{k-1}{N_\Psi-1}\,\frac{\pi}{2},
\qquad k=1,\dots,N_\Psi,
\]
spanning the full range from the flat state ($\Psi=0$) to the fully deployed state ($\Psi=\pi/2$) (where $N_{\Psi}$ are the number of frames required). For each deployment angle $\Psi_k$, the endpoints of every panel are updated using the same kinematic mapping as the popup mechanism (refer Eq.~\ref{eq:unit_kinematics}), yielding the deployed coordinates of the panel centrelines. The corresponding triangulated surface mesh is assembled and exported as an STL file. The output is therefore a sequence of STL meshes, encoding the complete deployment motion of the branching popup structure.

\subsection{2D cut-fold pattern generation}

\noindent To generate a fabrication-ready cut-fold pattern, the line-based CSV output is mapped onto a two-dimensional sheet. Each panel that makes up the popup unit centerline is converted into a strip element in the planar sheet representation. The vertical edges correspond to the cut boundaries, while the horizontal edges correspond to the fold (score/micro-cut) lines (refer Fig.~\ref{fig:Branching_structure}C). Unit-cell panels and support strips are treated consistently, and support strips contain fold lines only to preserve connectivity. The final 2D cut layout is exported as an SVG file, providing a resolution-independent vector representation that can be directly used for fabrication with digital cutting plotters.

\subsection{Fabrication procedure}

\noindent The SVG cut-fold pattern is then sent to Cricut Maker~3 cutting plotter and fabricated using 220~gsm sheets. The sheet is mounted on a standard-grip adhesive mat to prevent in-plane sliding during cutting. Cuts are performed with a fine-point blade with standard pressure settings appropriate for the selected material. After cutting, the sheet is removed and folded manually. Fold lines are implemented as scored dashed lines, which substantially reduce the folding effort and improve repeatability.

\subsection{Non-uniform unit-cell width}\label{subsec:width_variation}
\noindent In addition to controlling the discretization through the number of fold vertices $N$, our framework supports spatially varying unit-cell widths along the deployment direction. This is achieved by prescribing a slice-width vector $[w_1,\dots,w_{N_s}]$, which defines the local unit-cell width at each slice. Smaller values of $w_j$ can be assigned in high-curvature regions to increase the slice density, while larger values can be used in shallower regions to reduce the number of required unit-cells. Since the slice positions are computed by cumulative summation of $w_j$, modifying the width directly affects the slice-plane distribution, the optimized fold-vertex geometry, and the final branching cut-fold pattern. This provides a simple and flexible way to balance geometric fidelity and fabrication complexity. Finally, the framework enables modular design by stitching multiple optimized popup patches together.

In the prototype demonstrated in Fig.~\ref{fig:fig3}A, three independently designed surfaces corresponding to the positive, negative, and near-zero gaussian curvature regimes were fabricated and joined along their boundaries. For the composite spherical–hyperbolic–spherical configuration shown in Fig.~\ref{fig:fig3}A, the target surface is prescribed through a piecewise sinusoidal radius profile $\mathbf{r}(z)$, defining the embedding $\r^t (\theta,z)=\big(r(z)\cos\theta,\;1+r(z)\sin\theta,\;z\big).$ The radius function $r(z)$ is given by,
\begin{equation*}
r(z)
=
\begin{cases}
R_1 + (R_2 - R_1)\,\dfrac{1}{2}
\!\left[
1+\cos\!\left(\dfrac{\pi(z+L)}{L}\right)
\right],
& -2L \le z \le -L, \\[10pt]

R_2 + (R_3 - R_2)\,\dfrac{1}{2}
\!\left[
1+\sin\!\left(\dfrac{\pi z}{2L}\right)
\right],
& -L \le z \le L, \\[10pt]

R_3 + (R_2 - R_3)\,\dfrac{1}{2}
\!\left[
1-\cos\!\left(\dfrac{\pi(z-L)}{L}\right)
\right],
& L \le z \le 2L,
\end{cases}
\end{equation*}
where $R_1 > R_3 > R_2$. The first region exhibits the largest radius $R_1$ (positive curvature), the middle region attains the smallest radius $R_2$ (negative Gaussian curvature), and the final region transitions to an intermediate radius $R_3$, restoring the positive curvature with reduced magnitude. To accommodate curvature variations, the spatial discretization is prescribed as
\begin{equation*}
w(z)
=
\begin{cases}
\dfrac{L}{N}, & \text{Region 1 (largest radius)},\\[8pt]
\dfrac{L}{3N}, & \text{Region 2 (negative curvature)},\\[8pt]
\dfrac{L}{2N}, & \text{Region 3 (intermediate radius)}.
\end{cases}
\end{equation*}

\begin{table}[h]
\centering
\resizebox{0.3\textwidth}{!}{%
\begin{tabular}{|c|c|c|c|}
\hline
\textbf{Figure} & $N$ & $N_s$ & $w$ (cm) \\ 
\hline
Fig.~\ref{fig:fig1}A & 1 & 1  & 3 \\ 
\hline
Fig.~\ref{fig:fig2}A & 3 & 10  & 1 \\ 
\hline
Fig.~\ref{fig:fig2}B & 5 & 10 & 1 \\ 
\hline
\multirow{3}{*}{Fig.~\ref{fig:fig3}A}
& $R_1$ region - 1 & 11 & 1.2 \\ 
\cline{2-4}
& $R_2$ region - 2 & 26 & 0.4 \\ 
\cline{2-4}
& $R_3$ region - 3 & 15 & 0.8 \\ 
\hline
Fig.~\ref{fig:fig4}A & 7 & 1 & 2 \\ 
\hline
Fig.~\ref{fig:fig4}B & 4 & 7 & 1.5 \\ 
\hline
Fig.~\ref{fig:fig4}C & 3 & 5& 1.5 \\ 
\hline
\end{tabular}%
}
\caption{\textbf{\textsf{Experimental details of popup structures.}} The popup structures fabricated using the design pipeline discussed in the main text are made of 220 gsm paper. The structure in Fig.~\ref{fig:fig4}B alone is made with 150 gsm Polypropylene sheet. Here $N$ is the unit cells per slice, $N_s$ is the number of slices and $w$ is the unit-cell width.}
\label{tab:expdetails}
\end{table}

\begin{figure}[htbp]
\centering
\includegraphics[width=0.75\textwidth]{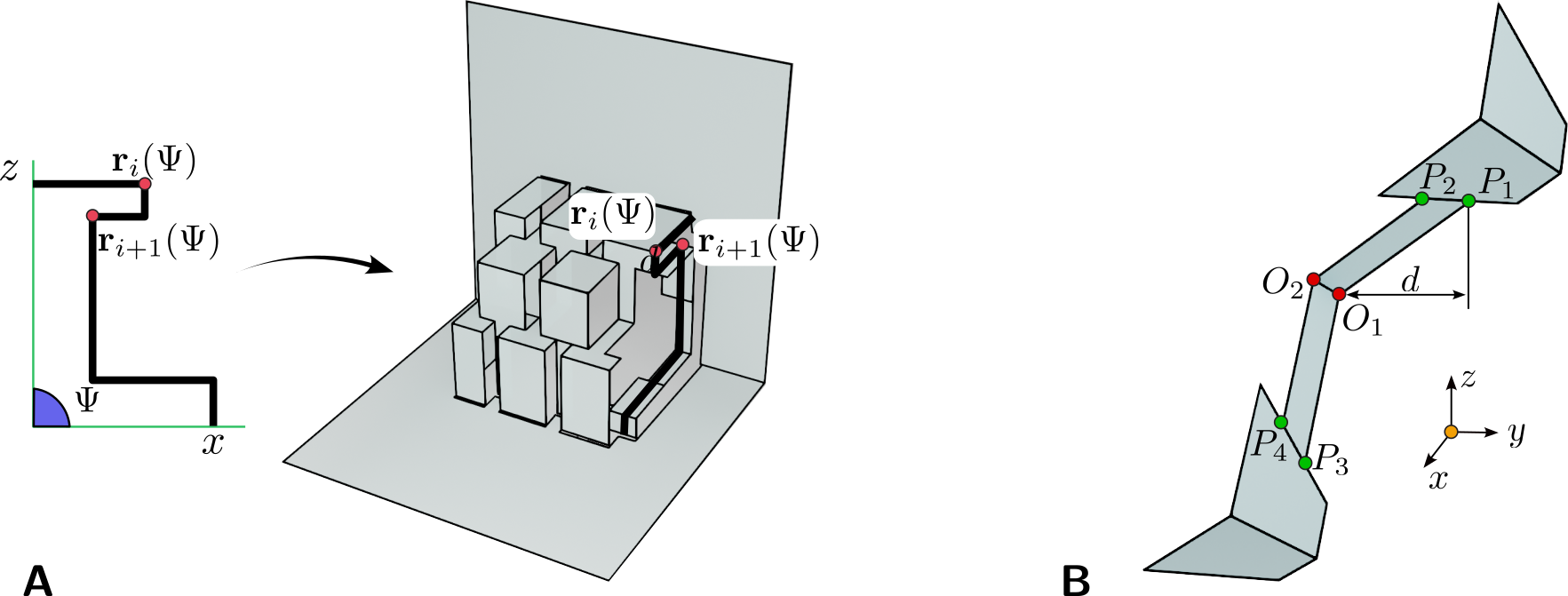}
\caption{\textbf{\textsf{Topological constraints in popup deployment.}}
(A) Illustration of a topological failure occurring when the deployed radial distance decreases along the sequence of units, i.e., $\r_{i+1}(\Psi) < \r_i(\Psi)$, leading to self-intersection and an infeasible configuration. (B) Schematic of two splayed popup units with connecting strips. Points $P_i$ (green) denote fixed edge points, while $O_1$ and $O_2$ (red) represent the fold-intersection points generated after actuation. Adjacent unit cells have opposite fold slopes $\alpha=\tan\gamma$ and are also actuated by the angle $\Psi$. The relative displacement between consecutive cells is represented by the offset $d$ between $P_1$ and $O_1$, which is determined by the inter unit-cell compatibility. The coordinate frame $(x,y,z)$ indicates the global reference used to compute the spatial positions of the points. See Sec.~\ref{subsection:connecting_strips} for more details.
}
\label{fig:SIfig3}
\end{figure}

\section{Geometry of splayed unit-cell}\label{SIsplayed}
In this section, we derive the expression for the angle
$\theta$ of the splayed popup unit. The unit-cell is defined by the four points $O$, $A$, $B$, and $C$, with $O$ as the spine and $A$ the fold vertex on the moving strip (refer Fig.~\ref{fig:fig3}C). The vectors $\r_{\rm OA}$, $\r_{\rm OB}$, and $\r_{\rm OC}$ lie in planes that make a prescribed slope with respect to $\r_{\rm OA}$, $\alpha = \tan\gamma$, where $\gamma$ is the inclination of the sloped folds. The actuation angle of the popup unit
is denoted by $\Psi$, and $\theta$ is the angle between $\r_{\rm OA}$ and the
$x$-axis in the base plane. The vector $\r_{\rm OA}$ is of length $w$ and it can be written as $\r_{\rm OA} = x\,\hat{\mathbf{e}}_x + y\,\hat{\mathbf{e}}_y$. Similarly, the point $B$ is obtained by following a fold of length $w \alpha$ inclined by $\gamma$ relative to $\r_{\rm OA}$ and rotated by the actuation angle $\Psi$. In the chosen coordinates, this can be written as,
\begin{align*}
\r_{\rm OB}
&= w\,\hat{\mathbf{e}}_x
 + w \alpha \cos (\Psi/2)\hat{\mathbf{e}}_y
 + w \alpha \sin (\Psi/2)\hat{\mathbf{e}}_z.
\end{align*}
By construction the triangle $OAB$ such that it is right-angled at $A$, we can use $\r_{\rm OA}\perp\r_{\rm AB}$ and obtain the condition,
\begin{align*}
x w - x^2 + y w \alpha \cos (\Psi/2) - y^2 &= 0.
\end{align*}
Using the length constraint $x^2 + y^2 = w^2$, we arrive at the trajectory of $\r_{\rm OA}$ as a function of $w, \alpha, \Psi$ as
\begin{align*}
x(\alpha,\Psi)
&= w\,\frac{1 - \bigl(\alpha\cos (\Psi/2)\bigr)^2}
           {1 + \bigl(\alpha\cos (\Psi/2)\bigr)^2}, \quad y(m,\Psi)
= w\,\frac{2 \alpha \cos (\Psi/2)}
           {1 + \bigl(\alpha\cos (\Psi/2)\bigr)^2}.
\end{align*}
The orientation of the vector $\r_{\rm OA}$, denoted by $\theta$, is given by
\begin{align}
\tan\theta(\alpha, \Psi)
&=
\frac{2 \alpha \cos (\Psi/2)}
     {1 - \big(\alpha\cos (\Psi/2) \bigr)^2}.
\end{align}

\subsection{Geometry of connecting strips between consecutive unit-cells}\label{subsection:connecting_strips}
\noindent We now determine the spatial coordinates of the strip that connects two consecutive splayed unit-cells (see Fig.~\ref{fig:SIfig3}B). The fixed points of the strip are denoted by $P_i$ (green dots), with coordinates $ P_i = (a_i,\, g_i,\, b_i),$ where $a_i$, $g_i$, and $b_i$ represent the $x$-, $y$-, and $z$-coordinates, respectively. The unknown fold-intersection points generated after actuation are denoted by $O_1$ and $O_2$ (red dots). The splay of the folds is characterized by the slope $\alpha = \tan\gamma$, and the mechanism is actuated by the angle $\Psi$. The coordinates of the red target point are obtained by enforcing fold alignment with the rotated unit-cell and inter-cell compatibility with the neighboring cell of opposite slope; since consecutive cells alternate the sign of $\alpha$, actuation induces a relative displacement that must be accounted to maintain continuity along the strip direction through the spacing difference $(g_2 - g_1)$. This relative displacement is represented by the offset $d$ shown in Fig.~\ref{fig:SIfig3} and is given by,
\begin{equation*}
d =
\frac{b_1 - b_3}{\tan\alpha}
\left[
1 -
\frac{1 - (\tan\alpha \cos(\Psi/2))^2}
     {1 + (\tan\alpha \cos(\Psi/2))^2}
\right].
\end{equation*}
Consequently, the $y$-coordinate of the red point $O_1$ is shifted relative to the corresponding edge point and is written as $ O_1 = (x,\, g_1-d,\, z),$ where the unknowns $x$ and $z$ determine its position in the $xz$-plane. The coordinates $(x,z)$ are obtained by enforcing two geometric constraints expressed through dot products of the fold vectors. Defining the vectors $\r_{\rm O_1P_1}, \r_{\rm P_1P_2}, \r_{\rm O_1P_3}, \r_{\rm P_3P_4},$ the fold-alignment condition is written as, 
$\r_{\rm O_1P_1} \cdot \r_{\rm P_1P_2}=\r_{\rm O_1P_3} \cdot \r_{\rm P_3P_4}$, ensuring consistent projection of $O_1$ along adjacent fold directions. The prescribed inclination of the splayed folds is enforced through the condition,
$\r_{\rm O_1P_1} \cdot \r_{\rm P_1P_2}=\cos\alpha$, which constrains the angle between the corresponding fold directions. These two linear equations in the unknowns $x$ and $z$ uniquely determine the spatial position of the red point $O_1$, completing the geometric construction of the connecting strip. An analogous construction applies to the red point $O_2$, whose coordinates are obtained by enforcing the same fold-alignment and inter-cell compatibility conditions with the corresponding adjacent edge points, thus fully determining the spatial configuration of the connecting strip.

\section{Popup design using first and second fundamental forms}\label{SI_augmented_loss}
\noindent In this section, we show that the design of popups can also be performed using the invariant representation of the surface connecting the fold vertices of the five-unit-sell assembly discussed in the main text. We do this by computing the discrete versions of the first fundamental form, $\mathbf{a}(r,\lambda)$ and the second fundamental form, $\mathbf{b}(r,\lambda)$ at the interior fold vertex $v_0(r,\lambda)$ (see SI Fig.~\ref{fig:curvature_illustration}). The computation uses the neighborhood consisting of the four adjacent vertices $v_1(\lambda), v_2, v_3(\lambda), v_4$, which define four incident triangles around $v_0$. Using $\mathbf{a}(r,\lambda)$ and $\mathbf{b}(r,\lambda)$ we can evaluate the appropriate combination of $r, \lambda$ that provides us with a popup structure with desired curvature.
\begin{figure}[htbp]
\centering
\includegraphics[width=\textwidth]{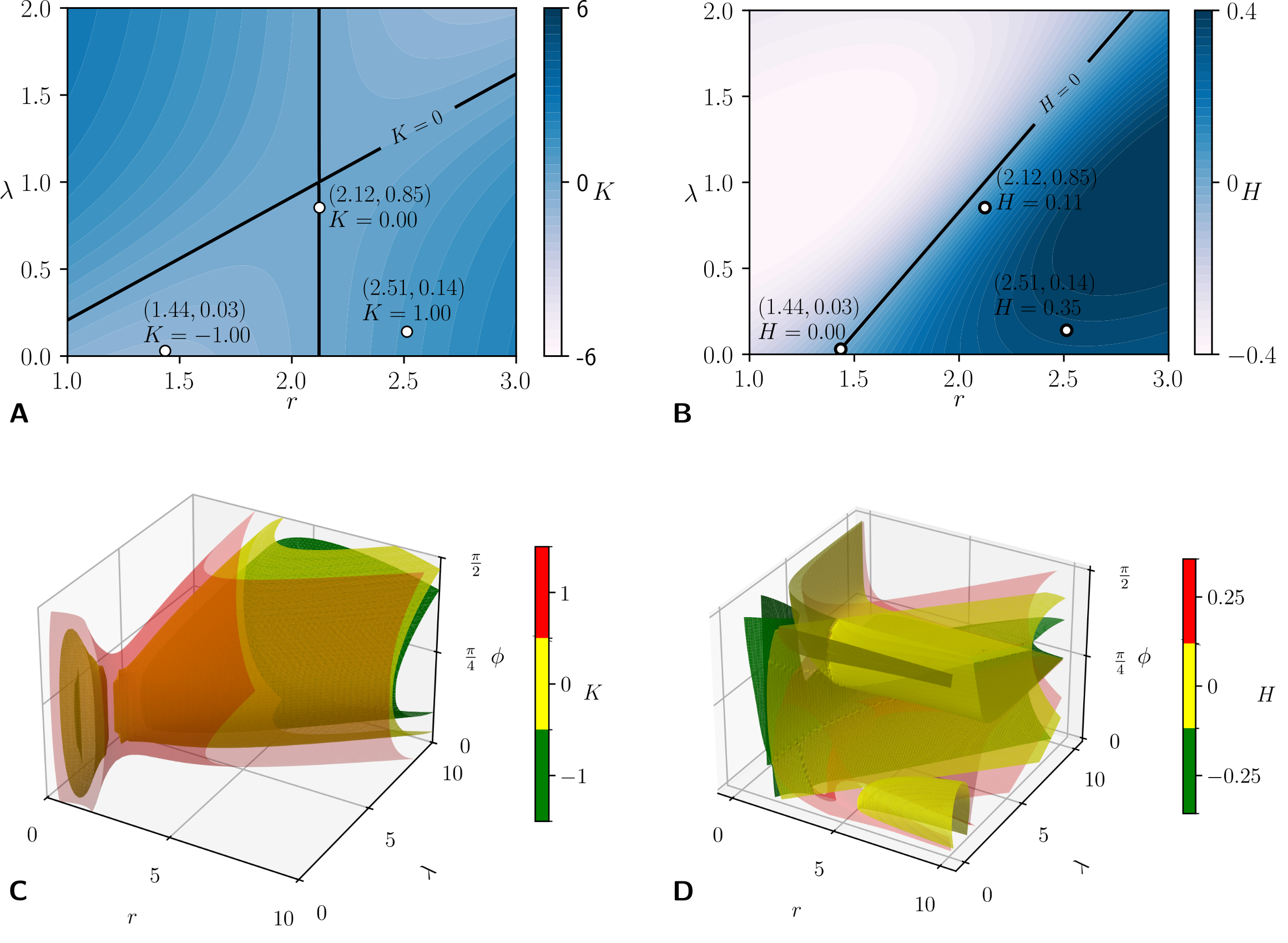}
\caption{
\textbf{\textsf{Curvature in the $(r,\lambda, \phi)$ parameter space.}} (A) Two-dimensional contour slice of the Gaussian curvature $K(r,\lambda)$ at $\phi=\pi/4$ (variables defined in Fig.~\ref{fig:fig1}). The black curve denotes the zero-Gaussian curvature locus $K=0$, separating elliptic and hyperbolic regions. The marked points are optimal $(r^*, \lambda^*)$ pairs obtained by minimizing the loss, $\mathcal{L}$ corresponding to minimizing curvature error and smoothness cost in Eq.~\ref{eq:optimization}. The labels indicate the coordinates and curvature values obtained from minimization. (B) Two-dimensional contour of the mean curvature $H(r,\lambda)$ at $\phi=\pi/4$. The black curve represents the minimal surface condition $H=0$. The highlighted points correspond to optimal solutions obtained by minimizing the augmented loss, $\L$. (C, D) Three-dimensional isosurfaces of the Gaussian curvature $K(r, \lambda, \phi)$ and the mean curvature $H(r, \lambda, \phi)$ over the full parameter domain. Green denotes $K<0$ (hyperbolic), $H<0$, yellow indicates $K=0$ (isometric), $H=0$ (minimal surface) and red represents $K>0$ (elliptic), $H>0$.}
\label{fig:SI_fig5_newloss}
\end{figure}
\subsection{Loss function definition}
To bias the deployed geometry towards a prescribed curvature, we introduce a penalty based on deviations of the computed fundamental forms of the target geometry, $(\tilde{\mathbf{a}}, \tilde{\mathbf{b}})$. The quadratic loss, $\mathcal{L}_{\mathrm{inv}}$ associated with this deviation is given by,
\begin{equation*}\label{eq:agumented_loss}
\mathcal{L}_{\mathrm{inv}}(r,\lambda)
=
\lambda_a \,\|\mathbf{a}(r,\lambda)-\tilde{\mathbf{a}}\|^2
+
\lambda_b \,\|\mathbf{b}(r,\lambda)-\tilde{\mathbf{b}}\|^2.
\end{equation*}
Now, we know that the Gaussian curvature $K(r,\lambda)$ and the mean curvature $H(r,\lambda)$ for the five-unit-cell assembly are uniquely determined by the first and second fundamental forms through the relations:
\begin{equation*}
K(r,\lambda)=\frac{\det\mathbf{b}(r,\lambda)}{\det \mathbf{a}(r,\lambda)},
\qquad
H(r,\lambda)=\frac{1}{2}\mathrm{tr}\!\left(\mathbf{a}(r,\lambda)^{-1}\mathbf{b}(r,\lambda)\right).
\end{equation*}
Therefore, penalizing deviations in $(\mathbf{a},\mathbf{b})$ is equivalent to directly enforcing the target curvature values $(\tilde{K},\tilde{H})$.

In the main article, we showed that the deployed five-unit-cell assembly can realize different curvature regimes depending on the design parameters $(r,\lambda)$. In particular, the Gaussian curvature field $K(r,\lambda)$ was computed using the discrete Gauss--Bonnet formulation (see SI Sec.~\ref{Gaussian-and-mean}) and visualized as a contour plot over (refer Fig.~\ref{fig:fig1}C) the $(r,\lambda)$-space. Representative designs for $K=-1$, $K=0$, and $K=1$ were originally obtained by minimizing the deployment loss $\mathcal{L}_j$ (ref. Eq.~\ref{eq:optimization}), where curvature emerged indirectly from the optimized geometry. Here we show that we can explicitly enforce curvature behavior through the loss, $\mathcal{L}_{\mathrm{inv}}(r,\lambda)$ or equivalently through a curvature-based loss written directly in terms of Gaussian and mean curvature:
\begin{equation*}
\mathcal{L}_{\mathrm{curv}}(r,\lambda)
=
\lambda_K (K(r,\lambda)-\tilde{K})^2
+
\lambda_H (H(r,\lambda)-\tilde{H})^2.
\end{equation*}
The complete objective is therefore $\mathcal{L} = \mathcal{L}_j(\{l_x^{i, j}, l_z^{i, j}\}) + \mathcal{L}_{\mathrm{curv}}(r,\lambda)$,
which is equivalent to the formulation based on fundamental-forms. The optimized parameter pairs, $(r^*, \lambda^*) = \arg \min_{r, \lambda} \mathcal{L}$ are shown in Fig.~\ref{fig:SI_fig5_newloss}A, B. The solutions obtained have gaussian curvatures, $K=-1$ (saddle), $K=0$ (developable), and $K=1$ (spherical). The gaussian and mean curvature values evaluated in these optimized configurations confirm that the augmented formulation enforces the desired curvature regimes while maintaining a uniform unit-cell deployment. This shows that the curvature classification previously obtained by the discrete curvature map $K(r,\lambda)$ can be achieved equivalently by incorporating curvature constraints directly into the optimization framework.
\end{document}